# First principles studies of the structural, electronic and optical properties of LiInSe$_2$ and LiInTe$_2$ chalcopyrite crystals

C.-G. Ma[a], M.G. Brik[a,b,c*]

[a] College of Sciences, Chongqing University of Posts and Telecommunications, Chongqing 400065, P.R. China

[b] Institute of Physics, University of Tartu, Ravila 14C, Tartu 50411, Estonia

[c] Institute of Physics, Jan Dlugosz University, Armii Krajowej 13/15, PL-42200 Czestochowa, Poland

**Abstract**

Detailed first principles calculations of the structural, electronic and optical properties of two representatives of the chalcopyrite group of compounds (LiInSe$_2$ and LiInTe$_2$) are reported in the present paper. Both materials are shown to be the direct band gap semiconductors. Analysis of the electronic properties showed that the degree of covalency increases if Se is substituted by Te. Calculations of the optical properties of both crystals allowed getting reliable approximation of the refractive index as a function of the wavelength. All calculated results were compared with the available experimental data; good agreement was demonstrated.

**Key words:** chalcopyrite; electronic structure; optical properties; ab initio calculations

## 1. Introduction

Chalcopyrite crystals with the general chemical formula I-III-VI$_2$ (e.g. CuGaS$_2$, CuInS$_2$ etc) [1,2,3] are used in numerous optical applications, such as solar cells, non-linear optical devices etc. They can be grown in a form of thin films, which is important for their applications in solar panels. Generally, most of these compounds have a rather narrow band gap, which – for many representatives of this group of materials – matches the maximum of solar spectrum. This is an additional argument in favor of their importance for solar energetics. Recently, remarkable success has been achieved in this area, resulting in a fast development of the Cu(In,Ga)(Se,S)$_2$-based solar elements, that already were advanced far enough to question a long dominance of the silicon-based solar panels [4]. Intensive research in this area has never stopped, and applications of new chalcopyrite materials – both neat and alloyed – are being reported. Thus, recently a high potential of LiInSe$_2$ (although in the orthorhombic phase) as a solar cell material was shown [5]. Theoretical calculations of physical properties of the LiInSe$_2$ polymorphs were reported in Ref. [6]; orthorhombic LiInS$_2$ and LiInSe$_2$ crystals were described in Ref. [7], their lattice dynamics and thermodynamic properties were calculated in Ref. [8], and elastic properties were reported in Ref. [9]. As far as LiInTe$_2$ is concerned, its electronic structure was studied in Ref. [10], whereas the phonon properties were calculated in Ref. [11].

---

[*] Corresponding author. E-mail: brik@fi.tartu.ee Phone: +372 7374751

In the present paper a comparative study of the structural, electronic, and optical properties of LiInSe$_2$ and LiInTe$_2$ crystals is performed. Special attention is paid to the role of the anions in the formation of the peculiar features of the electronic and optical properties of these compounds. There is a certain lack of the experimental and theoretical information on the optical properties of these crystals; the present article fills in this gap. Besides, a deeper insight into the nature of the chemical bonds in its relation to the electronic proeprties in both materials is gained.

The structure of the paper is as follows: in the next section the structure of the considered materials along with the computational method is described. The paper is continued with the description and discussion of the calculated electronic and optical properties and is concluded with a short summary.

## 2. Crystal structure and details of calculations

One unit cell of LiInSe$_2$ is shown in Fig. 1. This material as well as LiInTe$_2$ can crystallize in the chalcopyrite structure, the space group I-42d with four formula units per one unit cell. In this structure each atom has four nearest neighbors: each cation is coordinated by four selenium (tellurium) ions, whereas each selenium (tellurium) ion has two Li and two In ions as the nearest neighbors. The crystal lattice parameters – both experimental and calculated ones – are collected in Table 1.

The calculations were performed with the CASTEP module[12] of the Materials Studio package. The generalized gradient approximation (GGA) with the Perdew-Burke-Ernzerhof functional [[13]] and the local density approximation (LDA) with the Ceperley–Alder–Perdew–Zunger (CA–PZ) functional [[14], [15]] were used to treat the exchange-correlation effects. The plane wave basis set cut-off energy was 350 eV, the Monkhorst-Pack scheme *k*-point grid sampling was set as 5×5×3 *k*-points for the Brillouin zone (BZ). The convergence tolerance parameters were as follows: energy $5\times10^{-6}$ eV/atom, maximal force and stress 0.01 eV/Å and 0.02 GPa, respectively, and the maximal displacement $5\times10^{-4}$ Å. The explicitly considered electron configurations were $1s^22s^1$ for Li, $4d^{10}5s^25p^1$ for In, $4s^24p^4$ for Se, and $5s^25p^4$ for Te. The calculations were performed for a primitive cell.

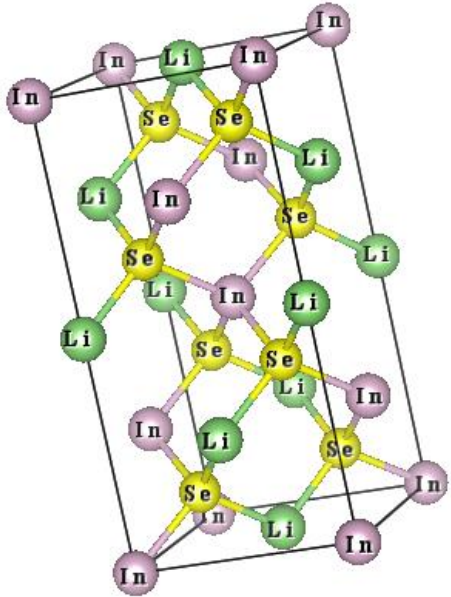

Fig. 1. One unit cell of LiInSe$_2$. Drawn using VESTA package [[16]].[16]

The choice of the GGA and LDA calculating techniques is justified by their successful application to modeling wide range of the structural, electronic, optical thermodynamic properties of various representatives of the chalcopyrite compounds reported very recently, e.g. CuGa(Se$_x$S$_{1-x}$)$_2$ [[17]], Cu*X*Te$_2$ (*X*=Al, Ga, In) [[18]], BeSi*V*$_2$ and MgSi*V*$_2$ (*V*=P, As, Sb) [[19]], ZnGeP$_2$ [[20]], ZnSiP$_2$ [[21]] etc.

Chalcopyrite semiconductors are known to have different kind of defects (off-stoichiometry, native and structural defects, as studied in CuInSe$_2$ [22]), which affect their properties. Consideration of such defects is beyond the scope of the present paper; all calculations were performed for the ideal structures. This is the first step towards analysis of the role played by the defects in the considered crystals.

**3. Results of calculations: structural and electronic properties**

The calculated values of the lattice constants of both considered crystals in comparison with the experimental results and other calculations (when available) are collected in Table 1. As seen from the Table, the GGA-calculated lattice parameters are somewhat greater than the LDA-calculated ones. It can be also noted that the LDA scheme gives better agreement with the experimental structural data, than the GGA.

Table 1. Summary of the experimental and theoretical lattice constants (all in Å) for the LiInSe$_2$ and LiInTe$_2$ compounds.

| Crystal | Experiment | | Calculated, this work | | | | Calculated | |
|---|---|---|---|---|---|---|---|---|
| | | | GGA | | LDA | | | |
| | $a$ | $c$ | $a$ | $c$ | $a$ | $c$ | $a$ | $c$ |
| LiInSe$_2$ | 5.807[a] | 11.810[a] | 6.0055 | 11.8754 | 5.8356 | 11.5937 | 5.818[c] | 11.53[c] |
| LiInTe$_2$ | 6.398[b] | 12.460[b] | 6.5462 | 12.8049 | 6.3000 | 12.4820 | 6.308[d] | 12.385[d] |

[a]Ref.23 [b]Ref.24 [c]Ref. [6] [d]Ref. [10]

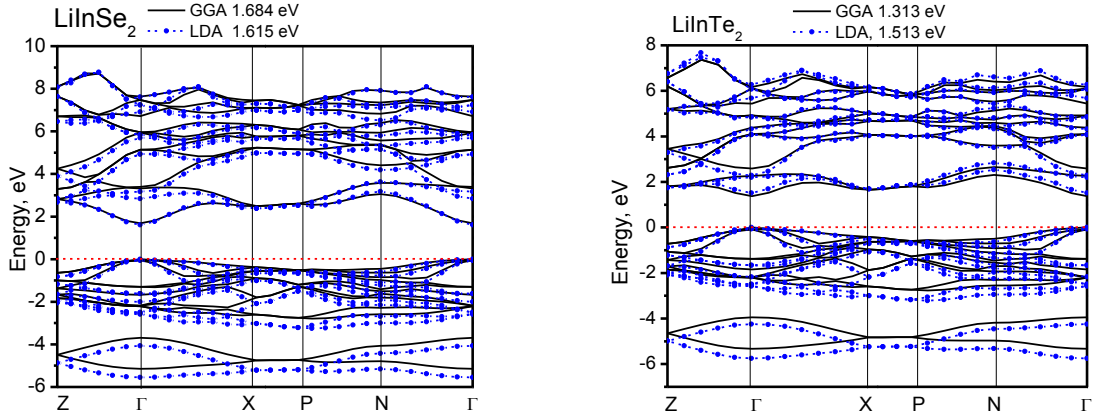

Fig. 2. Calculated band structures of LiInSe$_2$ and LiInTe$_2$. The coordinates of the special points of the Brillouin zone are (in units of the reciprocal lattice vectors): Z(1/2, 1/2, -1/2), Γ(0, 0, 0), X(0,0,1/2), P(1/4, 1/4, 1/4), N(0, 1/2, 0).

The calculated band structures of both materials are shown in Fig. 2. Both materials are the direct band gap semiconductors, with the calculated values (all in eV) were 1.684/1.615 (GGA/LDA) for LiInSe$_2$ and 1.313/1.513 (GGA/LDA) for LiInTe$_2$. Both values give the underestimated results comparing to the experimental data, which are 2.05 eV [23] for LiInSe$_2$ and 2.41 eV for LiInTe$_2$ [25]. Such an underestimation is not surprising for the density functional theory (DFT)-based methods; it is a standard practice to overcome it – if needed – by introducing the scissor operator, which produces a rigid up-ward shift of the conduction band until the band gap matches the experimental value. Such a scissor operator was not used in the present article. It can be also noted that the use of hybrid exchange–correlation DFT/Hartree–Fock (HF) scheme [26] improves the calculated band gaps.

    The dispersion of the electronic states is well pronounced in both conduction and valence bands around the Brillouin zone center (Γ point). The effective mass of electrons around the Γ point is lower in LiInSe$_2$ than in LiInTe$_2$, which follows from the comparison of the curvatures of the electronic states in the conduction band around the Γ point. At the same time, nearly all electronic states are remarkably flat along the X-P direction (Fig. 2), thus showing a very low mobility of the charge carriers along that line in the reciprocal space.

The calculated electronic bands can be assigned with the help of the density of states (DOS) diagrams shown in Fig. 3. The lower parts of the conduction band in both compounds are made predominantly of the 5p and 5s states of In, whereas the Li 2s states are responsible for the top of the conduction bands. The valence band is about 5 eV wide and consists of the 4p (5p) states of Se (Te), respectively. A lower valence band between -15 eV and -10 eV in $LiInSe_2$ has two well-distinguished peaks at about -14 eV (4d states of In) and -11 eV (4s states of Se). These two peaks become better separated in $LiInTe_2$, since the 5s states of Te are located somewhat higher in energy (at about -10 eV) than the 4s states of Se. Finally, the 1s states of Li produce a very deep and sharp peak at about -44 eV in $LiInSe_2$ and about -45 eV in $LiInTe_2$.

The calculated effective Mulliken charges of all ions are collected in Table 2. All ionic charges differ considerably from those, which could be expected from the chemical formula. However, the effective charge of Li ions is close to its formal charge +1, which indicates that the Li-Se(Te) bonds are more ionic than the In-Se(Te) bonds, which are highly covalent.

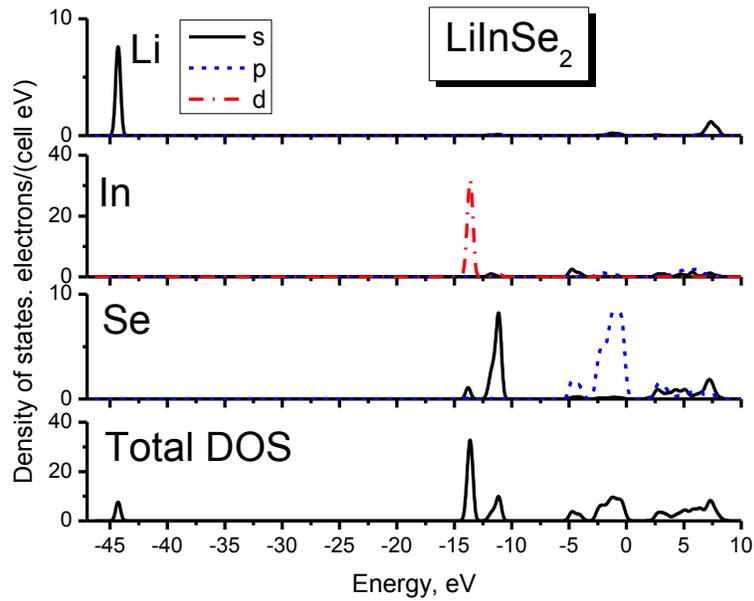

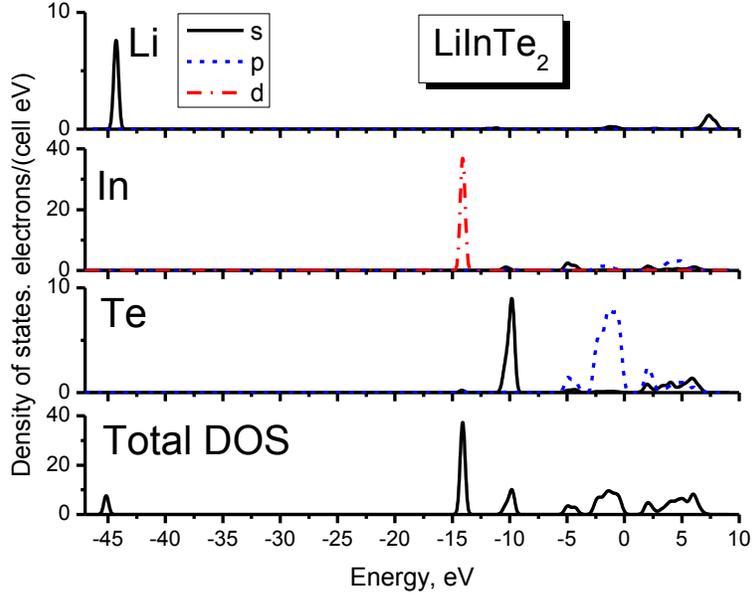

Fig. 3. Calculated density of states diagrams for LiInSe$_2$ and LiInTe$_2$.

The data from Table 2 are confirmed by the calculated cross-sections of the electron density difference (Fig. 4) in both LiInSe$_2$ and LiInTe$_2$.

Table 2. Calculated atomic populations for LiInSe$_2$ and LiInTe$_2$ (GGA/LDA values are given).

| Ion | s | p | d | Total | Charge (e) |
|---|---|---|---|---|---|
| LiInSe$_2$ | | | | | |
| Li | 2.20/2.18 | 0/0 | 0/0 | 2.20/2.18 | 0.80/0.82 |
| In | 1.43/1.44 | 1.42/1.52 | 9.99/9.99 | 12.84/12.96 | 0.16/0.04 |
| Se | 1.66/1.66 | 4.82/4.77 | 0/0 | 6.48/6.43 | -0.48/-0.43 |
| LiInTe$_2$ | | | | | |
| Li | 2.26/2.23 | 0/0 | 0/0 | 2.26/2.23 | 0.74/0.77 |
| In | 1.62/1.61 | 1.57/1.67 | 10.00/9.99 | 13.18/13.27 | -0.18/-0.27 |
| Te | 1.61/1.61 | 4.67/4.64 | 0/0 | 6.28/6.25 | -0.28/-0.25 |

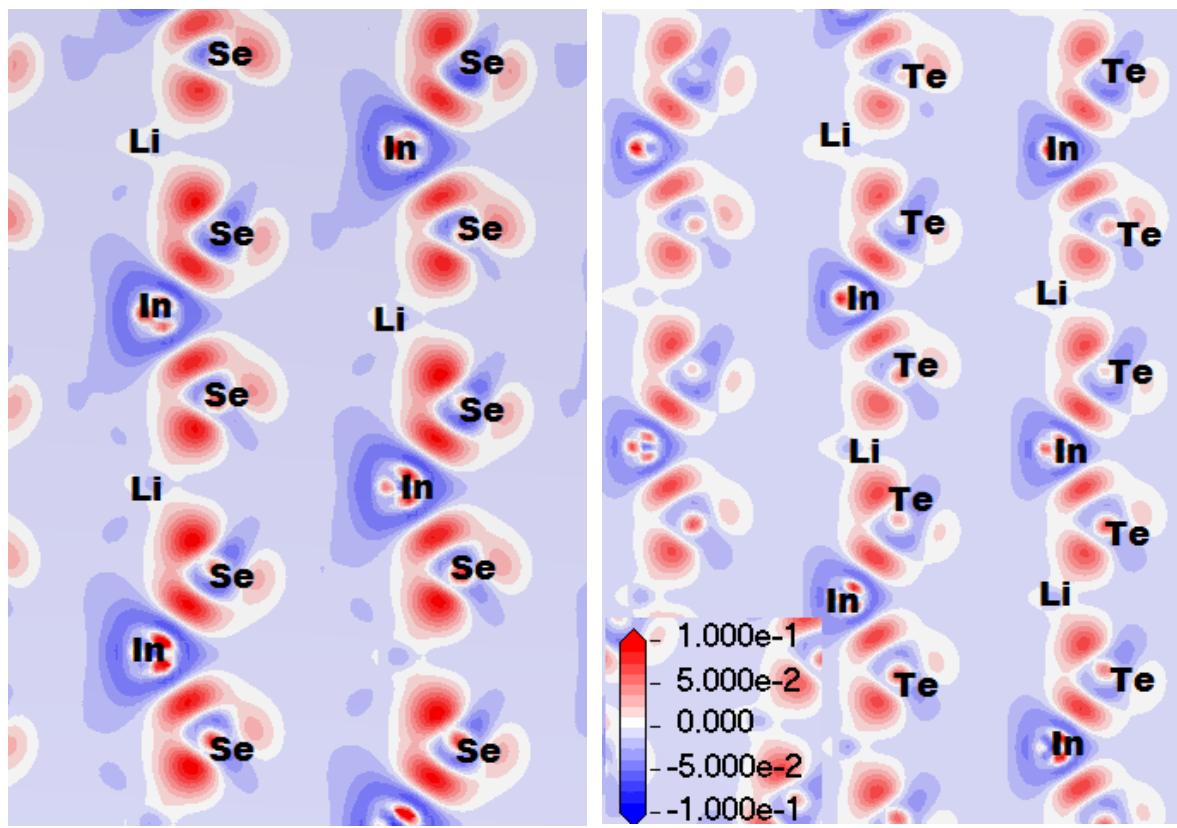

Fig. 4. Calculated cross-sections of the electron density difference in LiInSe$_2$ and LiInTe$_2$. The scale is in electrons/Å$^3$.

The areas, which gain electron density, are shown in red color in Fig. 4, whereas the areas loosing electron density are shown in blue color. The limits for the variation of the density difference are the same in both crystals: from -0.1 to +0.1 electrons/Å$^3$. The areas around Li ions appear to be close to white color – which means that the there is almost no difference between the electron density of a free ion of Li$^+$ and Li ions in both crystal lattices. This is also confirmed by the closeness of the effective and fomal charges of Li ions. The areas around Se ions are seen to gain more electron density than the areas around Te ions, which corresponds to the greater (in absolute value) charge of Se ions when compared to that of Te ions.

The difference of the calculated effective charges of two neighboring ions can serve as a measure for the degree of covalency of the chemical nond between them. The greater such difference is, the more ionic is the chemical bond. Apparently, the In-Te bond is more covalent than the In-Se (see Table 2 with effective charges of all ions).

## 4. Results of calculations: optical properties

After the crystal structure has been optimized and the wave functions of the calculated electronic states were obtained in a numerical form, one can proceed with the analysis of the optical properties by calculating the dielectric function of a solid. The imaginary part Im($\varepsilon(\omega)$) of a dielectric function $\varepsilon(\omega)$ (directly related to the absorption spectrum of a solid) is evaluated by numerical integrations of matrix elements of the electric dipole operator between the occupied states $\Psi_k^v$ in the valence band and empty states $\Psi_k^c$ in the conduction band:

$$\mathrm{Im}(\varepsilon(\omega)) = \frac{2e^2\pi}{\omega\varepsilon_0} \sum_{k,v,c} \left|\left\langle \Psi_k^c \left| \vec{u} \cdot \vec{r} \right| \Psi_k^v \right\rangle\right|^2 \delta\left(E_k^c - E_k^v - E\right), \qquad (1)$$

where $\vec{u}$ is the polarization vector of the incident electric field, $\vec{r}$ and $e$ stand for the electron's position vector and electric charge, respectively, $E = \hbar\omega$ is the incident photon's energy, and $\varepsilon_0$ is the vacuum dielectric permittivity. The summation in Eq. (1) is performed over all states from the occupied and empty bands.

The real part Re($\varepsilon(\omega)$) of the dielectric function $\varepsilon(\omega)$, which determines the dispersion properties and refractive index values, is estimated in the next step by using the Kramers-Kronig relation:

$$\mathrm{Re}(\varepsilon(\omega)) = 1 + \frac{2}{\pi} \int_0^\infty \frac{\mathrm{Im}(\varepsilon(\omega'))\omega' d\omega'}{\omega'^2 - \omega^2} \qquad (2)$$

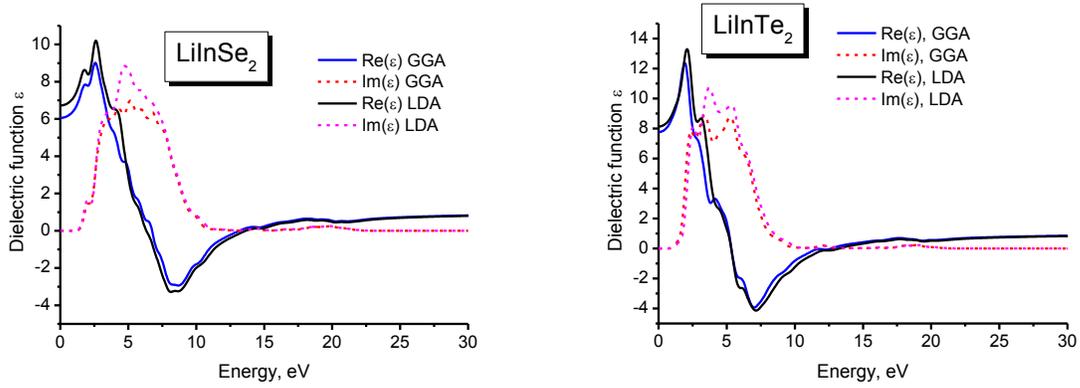

Fig. 5. Calculated real and imaginary parts of dielectric function for LiInSe$_2$ and LiInTe$_2$.

The calculated real and imaginary parts of the dielectric function for LiInSe$_2$ and LiInTe$_2$ are shown in Fig. 5; a minor difference between the GGA and LDA results can be seen, that is due to a difference in the calculated band gaps obtained in both approaches. A very broad absorption band (clearly seen in the Im(ε) plot) is related to the band-to-band absorption, or to the transitions between the occupied 4p (5p) states of Se(Te) in the valence band to the unoccupied 2s states of Li in the conduction band. Taking the square root from the values of $\mathrm{Re}(\varepsilon(\omega))$ in the limit of zeroth energy, the value of the refractive index $n$ can be evaluated, which are then 2.59 (LDA) and 2.46 (GGA) for LiInSe$_2$; the corresponding data for LiInTe$_2$ are 2.85 (LDA) and 2.79 (GGA). These results are in favorable agreement with the data reported in refs. [9, 27] for LiInSe$_2$ (2.48) and LiInTe$_2$ (2.81).

Fig. 6 shows the calculated dispersion of the refractive index $n$ in the spectral range corresponding to the normal dispersion – below the absorption edge. The wavelength dependence of the refractive index $n$ was fitted to the Sellmeier equation with the infrared correction [28]

$$n = A + \frac{B}{1-\left(\frac{C}{\lambda}\right)^2} - D\lambda^2 . \qquad (3)$$

The values of the $A$, $B$, $C$, $D$ coefficients extracted from such a fit are collected in Table 3. As Fig. 6 shows, the quality of fit is very good with the correlation coefficient of about 0.99.

Table 3. Parameters of the Sellmeier equation (3)

| Parameter | LiInSe$_2$ | | LiInTe$_2$ | |
|---|---|---|---|---|
| | GGA | LDA | GGA | LDA |
| $A$ | -1.55888 | -0.61826 | -0.68804 | 0.66561 |
| $B$ | 3.99384 | 3.19346 | 3.44426 | 2.16817 |
| $C$, nm | 196.74201 | 225.45098 | 277.77357 | 315.96737 |
| $D$, nm$^{-2}$ | $-5.02012 \times 10^{-10}$ | $-4.60871 \times 10^{-10}$ | $-6.47803 \times 10^{-10}$ | $-4.63767 \times 10^{-10}$ |

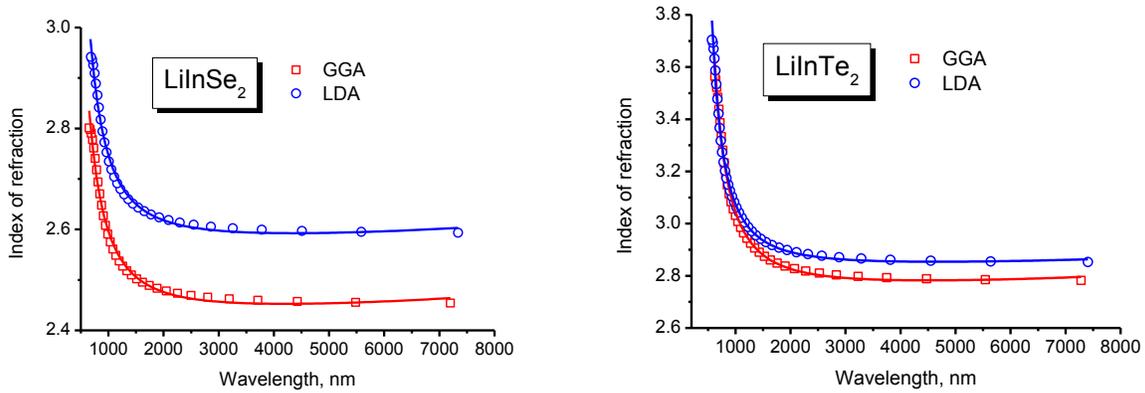

Fig. 6. Calculated refractive index (symbols) as a function of the wavelength for LiInSe$_2$ and LiInTe$_2$. The solid lines are the fits to Eq. (3).

With the help of Eq. (3) and the parameters from Table 3, one estimate the refractive index for both materials in the considered range of the wavelengths.

## 5. Conclusions

The structural, electronic and optical properties of two chalcopyrite semiconductors LiInSe$_2$ and LiInTe$_2$ were calculated in the present paper using the first principles calculations method (the CASTEP module of Materials Studio). Both compounds are the direct band gap semiconductors, with the maximum of the valence band and the minimum of the conduction band realized at the center of the Brillouin zone. In both compounds the Li ions appear to be bonded to the crystal lattices by the ionic bonds predominantly, whereas the covalency is strong in the In-Se and In-Te (especially) pairs. The calculated values of the dielectric function allowed considering the absorption properties of both materials (the optical absorption is due to the valence-to-conduction bands transitions) and dispersion of the refractive index. The latter was fitted to the Sellmeier equation. The parameters of fit allow for a reliable estimation of the refractive indices of both materials in a wide spectra range from ca. 600 nm to 8000 nm.

**Acknowledgment**

M.G. Brik appreciates financial support from Ministry of Education and Research of Estonia, project PUT430, the European Social Fund's Doctoral Studies and Internationalisation Programme DoRa, and the Programme for the Foreign Experts offered by Chongqing University of Posts and Telecommunications. C.-G. Ma thanks the financial support of National Natural Science Foundation of China (Grant No.11204393), Natural Science Foundation Project of Chongqing(Grant No. CSTC2014JCYJA50034) and Scientific and Technological Research Program of Chongqing Municipal Education Commission (Grant No. KJ110515). Dr. G.A. Kumar (University of Texas at San Antonio) is thanked for allowing us to use the Materials Studio package.